\documentclass[notitlepage,twocolumn,prx,superscriptaddress]{revtex4-2}
\usepackage{amsmath}
\usepackage{graphicx,placeins}
\usepackage[space]{grffile} %Import files with spaces in names
\usepackage{gensymb}
\usepackage{bm}% bold math
\usepackage{amssymb,amsfonts,amsmath,amsbsy,bm,t1enc,latexsym}
\usepackage{soul}
\usepackage{amssymb} %maths
\usepackage{amsmath} %maths
\usepackage{amsfonts}
\usepackage{amsmath}
\usepackage{amssymb}
\usepackage{upgreek}
\usepackage{graphicx}
\RequirePackage{lineno}
\usepackage{siunitx}
\usepackage{hyperref}       % hyperlinks
\hypersetup{
    colorlinks=true,
    linkcolor=blue,
    filecolor=magenta,      
    urlcolor=cyan,
}

\usepackage[defaultcolor=red]{changes}
\definechangesauthor[color=blue]{MA}

\usepackage[utf8]{inputenc} %useful to type directly diacritic characters

\newcommand*\SiN{Si$_3$N$_4$ }

\begin{document}
\title{Dissipative solitons and Switching waves in dispersion folded Kerr cavities}

\author{Miles H. Anderson}
\affiliation{Institute of Physics, Swiss Federal Institute of Technology (EPFL), Lausanne, Switzerland}

\author{Alexey Tikan}
\affiliation{Institute of Physics, Swiss Federal Institute of Technology (EPFL), Lausanne, Switzerland}

\author{Aleksandr Tusnin}
\affiliation{Institute of Physics, Swiss Federal Institute of Technology (EPFL), Lausanne, Switzerland}

\author{\\Johann Riemensberger}
\affiliation{Institute of Physics, Swiss Federal Institute of Technology (EPFL), Lausanne, Switzerland}

\author{Rui Ning Wang}
\affiliation{Institute of Physics, Swiss Federal Institute of Technology (EPFL), Lausanne, Switzerland}

\author{Tobias J. Kippenberg}
\email{tobias.kippenberg@epfl.ch}
\affiliation{Institute of Physics, Swiss Federal Institute of Technology (EPFL), Lausanne, Switzerland}

\date{\today}

\begin{abstract}
We theoretically and experimentally investigate the formation of dissipative coherent structures in Kerr nonlinear optical microresonators, whose integrated dispersion exceeds the free-spectral range. We demonstrate that the presence of any periodic modulation along the resonator's circumference, such as periodically varying dispersion, can excite higher-order comb structures. We explore the outcomes of coherent microcomb generation in both cases of anomalous and normal dispersion. We are able to access this regime in microresonators via the high peak power of synchronous pulse-driving.
For solitons in anomalous dispersion, we observe the formation of higher-order phase-matched dispersive waves (`Kelly-like' sidebands), where the folded dispersion crosses the frequency comb grid.
In normal dispersion, we see the coexistence of switching wave fronts with Faraday instability-induced period-doubling patterns, manifesting as powerful satellite microcombs highly separated either side of the core microcomb while sharing the same repetition rate. 
This regime of dispersion-modulated phase matching opens a dimension of Kerr cavity physics and microcomb generation, particularly for the spectral extension and tuneability of microcombs in normal dispersion. 
\end{abstract}

\pacs{}

%\tableofcontents

\maketitle

\section{Introduction}

\begin{figure*} [t]
	\centering
	\includegraphics{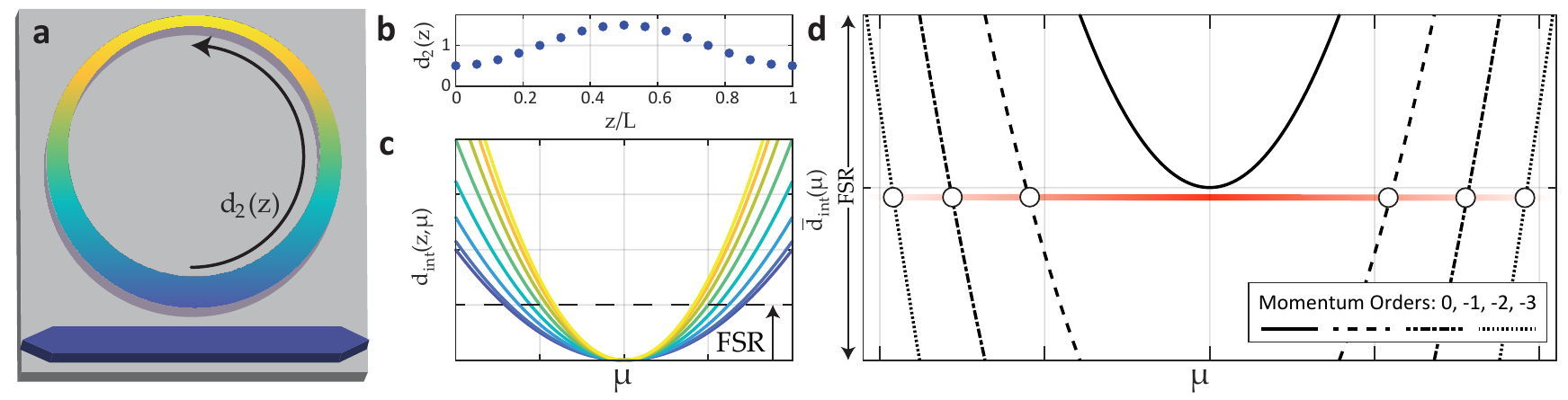}
	\caption{\textbf{Fundamental model under consideration}.(a) The simplest case of a passive resonator whose waveguide width, and thus dispersion parameter, undergoes a periodic cycle. (b) In our modelling we consider a simple sinusoidal dispersion modulation, evaluated piece-wise. (c) The consequent roundtrip excursion of the integrated dispersion of each resonator mode $\mu$, for each value of $d_2(z)$. (d) Roundtrip integrated dispersion, this time plotted in the folded zone over 1 free-spectral range. The comb spectrum spans multiple orders of dispersion, but each higher order is phase-mismatched. Dispersion modulation is required to quasi-phase match. }
	\label{fig:gentheory_concept}
\end{figure*}

Nonlinear pattern formation is a fascinating phenomena and ubiquitous in nature~\cite{nicolis1977self}. Over the past decade, a wide variety of dissipative coherent structures in continuous wave driven Kerr optical nonlinear microresonators has been observed and studied~\cite{Kippenberg2018, chang_integrated_2022}. It is now well understood that two fundamental localized dissipative structures can be generated via parametric interactions in microresonators: dissipative Kerr solitons (DKSs) in anomalous dispersion\cite{Herr2014Temporal}, and switching waves (SWs)~\cite{Xue2015Mode-lockedMicroresonators, anderson2020zero}(otherwise termed as platicons or `dark pulses' in certain configurations) in the normal dispersion regime, which in the CW-driven case are excited by mode crossing induced dispersion changes~\cite{xue2015NormaldispersionMicrocombsEnabled}. Both structures are governed by the Lugiato-Levefer (LLE) equation and constitute coherent optical frequency combs in the frequency domain. Such `microcombs', DKS-based microcombs in particular, have expanded the domain of optical frequency combs and have been used in numerous system level applications, due to their high repetition rates in the microwave domain and their compact chip-integrated form factor, into various application domains; including neuromorphic computing~\cite{Feldmann2021ParallelCore}, ultrafast ranging and parallel coherent LIDARs~\cite{Trocha2018UltrafastCombs,Riemensberger2020}, coherent telecommunications~\cite{Marin-Palomo2017}, astronomical spectrometer calibration~\cite{Obrzud2019AAstrocomb}, frequency synthesizers~\cite{Spencer2018AnPhotonics}, and atomic clock architectures~\cite{Newman2019ArchitectureClock}.
Given the relative maturity of research into the physics of dissipative Kerr structures, greater attention has been given to exploring physics and microcomb generation in non-trivial regimes. These regimes include alternative pumping schemes such as pulse-driving\cite{Obrzud2017TemporalPulses} and self-injection locked lasers\cite{Raja2019ElectricallyMicrocomb, stern2018battery}, and the use of complex dispersion profiles to extend the comb bandwidth of solitons with dispersive waves (DWs)~\cite{spencer_optical-frequency_2018, moille_ultra-broadband_2021}. Complex resonator structures are now the object of attention, including in particular resonators with integrated Bragg or photonic-crystal elements\cite{yu_spontaneous_2021}, and dual coupled-ring photonic `dimers'\cite{Tikan2021EmergentDimer} or photonic `molecules'\cite{helgason_dissipative_2021}.

Yet, in nearly all prior experimental studies of driven nonlinear optical microresonators, the Kerr frequency shift has been, heuristically, less than the free spectral range (FSR). 
In other words, the spectral extent of the generated coherent dissipative structure exhibits an integrated dispersion that is smaller than the free spectral range. The is predominantly the natural regime for typical microresonators with a large FSR over $\sim$50 GHz. Here we study the nonlinear dynamics beyond this realm. 
Using pulsed optical pumping~\cite{Obrzud2017TemporalPulses, Anderson2021PhotonicSolitons} of low repetition rate optical microresonators, we access the regime where the `dissipative structures' cover a bandwidth whose integrated dispersion exceeds the FSR of the resonator. Akin to electrons in periodic bands that gives rise to the Brillouin zone, we demonstrate how dispersion folding can occur. Specifically, when the dispersion folds back to the zone spanned by the FSR, we show that any periodic perturbation of the soliton during its roundtrip (e.g. induced via spatially varying dispersion) enables quasi-phase matching for emergent higher-order structures at the edge of the dispersion `zone'. In this way, periodic forcing of the cavity field every roundtrip via the dispersion constitutes a form of parametric driving. 
For solitons, this gives rise to higher-order dispersive waves~\cite{luo_resonant_2015, Nielsen2018Invited}, also identified as `Kelly sidebands' historically discovered in systems with periodic amplification and later soliton fiber lasers~\cite{kelly1992characteristic, peng2018build}.  Additionally, in CW-driven systems, dispersion modulation, or in fact parametric modulation of any system parameter, has long been known to lead to Faraday Instability (FI). FI patterns were most originally studied in vertically shaken fluid basins\cite{noauthor_stability_1954} and were observed to occur oscillating first at half the forcing frequency. From a general point of view, such dynamics in optical resonators are governed by partial differential equations with periodic coefficients. The associated field of study is called Floquet theory. Floquet dynamics and consequent optical FI has been discussed earlier mostly in the context of fiber-based devices~\cite{coen_modulational_1997, staliunas_parametric_2013, Conforti2014Modulational, Mussot2018ModulationFibers} operating in the quasi-CW regime where period-doubling dynamics~\cite{Bessin2019Real} as well as the competition between Turing and Faraday instability~\cite{Copie2016Competing} have been observed. The very same dynamics have also been studied for Bose-Einstein condensates \cite{zhu_parametric_2019}. 

Here, we provide experimental observation of higher order (5th) dispersive waves bound to a dissipative Kerr soliton microcomb, and the generation of strong satellite combs in a switching wave microcomb, powered by FI pattern formation not yet seen or studied in photonic microresonators. In the latter, this results in a 5$\times$ extension of the total comb bandwidth, but where the satellite combs share the same comb line spacing, but a different offset frequency. For both cases, we theoretically analyse the the dynamics behind the periodically perturbed LLE using the paradigm of 2D four-wave mixing (FWM) in the unfolded dispersion space. Thus, our results enrich the conventional thinking about dispersion by introducing and studying the consequence of dispersion folding.

\section{General model of dispersion modulated cavity}

The fundamental model underpinning all the phenomena presented and discussed in this work is depicted in Fig.~\ref{fig:gentheory_concept}. The resonator can be represented as a waveguide ring (Fig.~\ref{fig:gentheory_concept}(a)) whose cross-section varies in such a way that the group velocity dispersion parameter $d_2(z)$ ($\beta_2$ or $D_2$ as it is in the experimental sample) varies periodically over length $L$ and amplitude $\Delta$, or in the time domain as $t=z/D_1R$ for resonator with radius $R$ and free-spectral range (FSR) $D_1$. 
The system can be be evaluated piece-wise as in Fig.~\ref{fig:gentheory_concept}(b). Expanded over the comb frequency mode index $\mu$, the dispersion operator $d_\mathrm{int} =d_2\mu^2$ sweeps up and down over one roundtrip (Fig.~\ref{fig:gentheory_concept}(c)). The concept of dispersion folding is depicted in Fig.~\ref{fig:gentheory_concept}(d). The dispersion curve passing FSR/2 is folded back to -FRS/2. In this picture, momentum mismatch between the branches of dispersion is not compensated in non-modulated cavities. Thus, FWM interactions with folded modes become resonant only when there in a mechanism coupling two neighbouring FSRs.

To establish a clear line of reasoning and understand the dynamics of this system, we revisit and extend conclusions presented in previous studies~\cite{Conforti2014Modulational, Copie2016Competing, Mussot2018ModulationFibers, Bessin2019Real} looking at them from a different point of view that employs the notion of two-dimensional four-wave mixing (2D FWM).

\subsection{Model}

To model the nonlinear dynamics of the cavity with periodically modulated dispersion, we use a well-known form of the Lugiato-Lefever equation (LLE) with a time-dependent dispersion term~\cite{Conforti2014Modulational}. In the dimensionless units, the equation takes the form:

\begin{equation}\label{eq:LLE1}
    \frac{\partial \Psi}{\partial t} = -(1+i\zeta_0)\Psi + i [d_2^{(0)}+d_2(t)]\frac{\partial^2 \Psi}{\partial \varphi^2} + i|\Psi|^2\Psi + f(\varphi),
\end{equation}
where $\Psi(\varphi,t)$ describes the slowly-varying envelope of the optical field in the microresonator, $f(\varphi)$ the driving function (which may be a pulse profile), $\varphi$ is the azimuthal coordinate inside the cavity in the frame moving with velocity $d_1 = 2D_1/\kappa$ with $D_1=2\pi\cdot\mathrm{FSR}$, $\zeta_0 = 2\delta\omega_0/\kappa$ is the normalized laser-cavity detuning, $\kappa=\kappa_0+\kappa_\mathrm{ex}$ is the total linewidth of the resonator with internal linewidth $\kappa_0$ and coupling to the bus waveguide $\kappa_\mathrm{ex}$. Dispersion coefficients $d_2^{(0)}$, $d_2(t)$ and time $t$ are normalized on the photon lifetime so that $d_2=D_2/\kappa$ and $t=t'\kappa/2$ for real lab time $t'$. 
In this model, we denote $d_2^{(0)}$ as the averaged resonator dispersion with periodic modulation $d_2(t+T) = d_2(t)$, where $T =T'\kappa/2 =\pi\kappa/D_1$ is the normalized roundtrip time. If the driving function $f(\varphi,t)$ is also periodic in time with period $T$, we can assume that the field $\Psi$ has the same symmetry $\Psi(t+T) = \Psi(t)$, we can employ the Fourier transform
\begin{align}
  d_2(t) &= \sum_n \tilde{d}_{2}^{(n)} e^{-i d_1 n t}, \label{eq:Fourier_d2}\\ 
  \Psi(\varphi,t) &= \sum_{n\mu} \tilde{\psi}_{n \mu} e^{i \mu \varphi-i d_1 n t} \label{eq:Fourier_field}  
\end{align}
 and obtain an effective two-dimensional equation governing the Floquet dynamics (here $f$ is taken constant for simplicity, but the equation can be readily generalized):
\begin{align}\label{eq:2D_eq}
    & \frac{\partial \tilde{\psi}_{n \mu}}{\partial t} = -(1+i [\zeta_0-n d_1] + id^{(0)}_2 \mu^2)\tilde{\psi}_{n \mu} - \nonumber\\
    &-i \sum_{m} \tilde{d}_{2}^{(n-m)} \mu^2 \tilde{\psi}_{m \mu}     
    +i \sum_{\substack{n_1,n_2,n_3 \\ \mu_1,\mu_2,\mu_3}} \tilde{\psi}_{n_1 \mu_1}\tilde{\psi}_{n_2 \mu_2}\tilde{\psi}_{n_3 \mu_3}^* \delta_{\mathrm{FWM}}\nonumber \\
    &+\delta_{n,0}f,
\end{align}
where the conservation law $\delta_\mathrm{FWM} = \delta(\mu_1+\mu_2-\mu_3-\mu)\delta(n_1+n_2-n_3-n)$ governs 2D FWM processes in the fast($\mu$)- slow($n$)- frequency space. We can thus conclude that the periodically varying dispersion, which in the LLE leads to a time dependent dispersion term, couples different Floquet orders ($n$) of the intracavity field. 

\begin{figure} [t]
	\centering
	\includegraphics{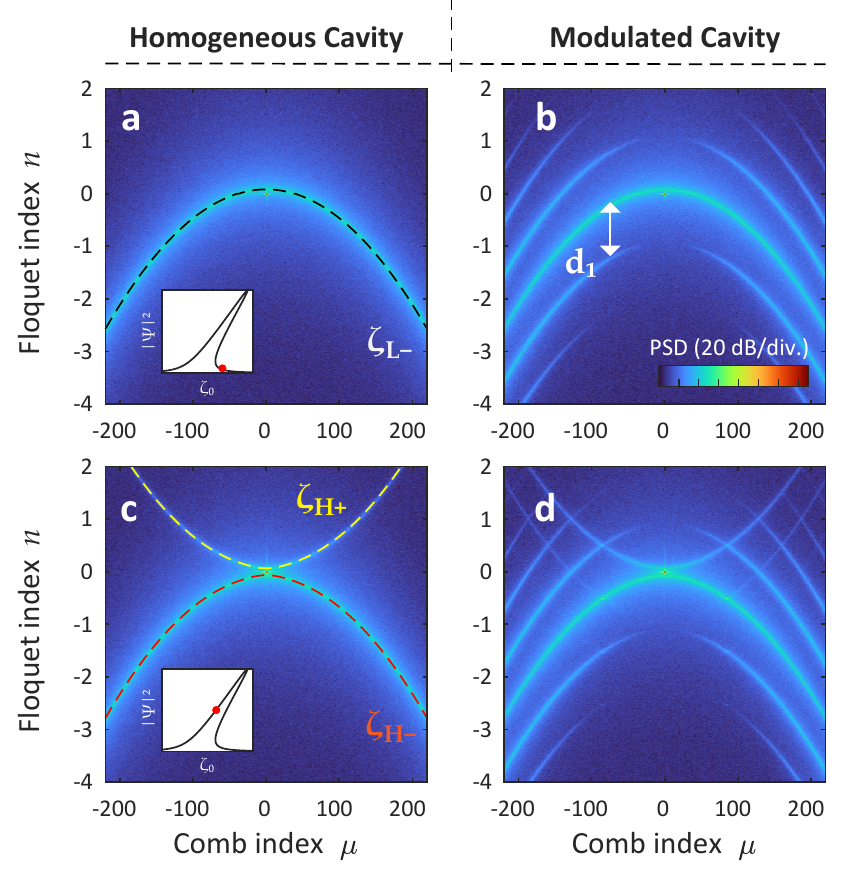}
	\caption{\textbf{Dispersion curves or Floquet bands for homogeneous and modulated cavity}. (a,c) Simulated 2D Fourier-transform image for light propagating in the Kerr lower-state at $\zeta_0=5$, and upper-state at $\zeta_0=4$ respectively, in homogeneous cavity with $f^2=10$. The conjugated dispersion curve is only apparent in the upper state. (b,d) Same solution existing in dispersion-modulated cavity with $\Delta=0.5 d_2^{(0)}$, $d_1/2\pi=8$, and $d_2^{(0)}=0.0027$.}
	\label{fig:cw_dispersion}
\end{figure}

\begin{figure*}[t]
	\centering
	\includegraphics[width=\linewidth]{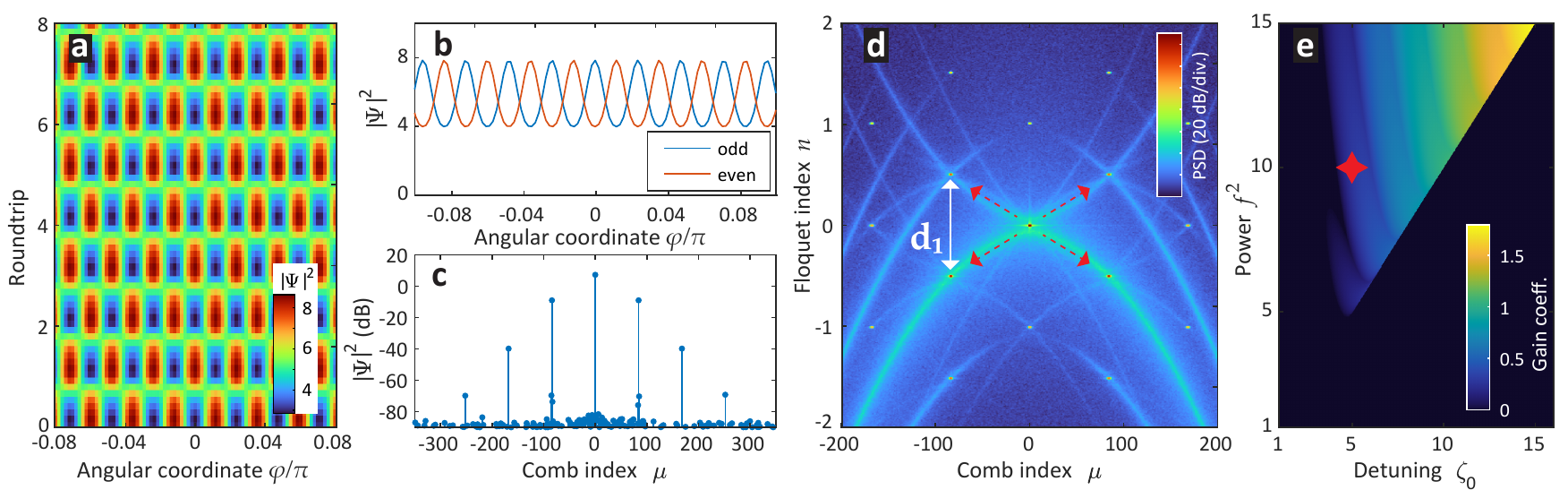}
	\caption{\textbf{Faraday instability (FI) simulation in CW-driven Kerr cavity with normal dispersion}. For cavity with $f^2=10$, $\zeta_0=5$, $\Delta=0.5 d_2^{(0)}$, $d_1/2\pi=8$, and $d_2^{(0)}=0.0027$. (a) Fluctuating field over roundtrips vs. angular coordinate. (b) Angular coordinate domain showing two consecutive cavity roundtrips. (c) Fast frequency domain snapshot. (d) 2D Fourier-transform (NDR) of (a) for Floquet mode index $n$ vs. comb mode index $\mu$. (e) Gain coefficient for FI as a function of detuning and power. Red star marks operating point for (a-d).}
	\label{fig:FI_CW}
\end{figure*}

The dispersionless profile along $n$ (modes are equidistant in this direction having $D_1$ frequency spacing) protects our system from transverse instabilities~\cite{Ablowitz2021Transverse, zakharov1973instability}, allowing us to study and generalize well-known coherent structures such as DKS and SW. However, this modulation results in linear coupling between different orders of Floquet index $n$ that effectively correspond to FSR-frequency breathing. As shown in Eq.~(\ref{eq:2D_eq}), the
coupling amplitude is proportional to the Fourier coefficients $\tilde{d}_2^{(n)}$ and scales \emph{quadratically} with comb index $\mu$, increasingly strengthening the coupling rate for larger mode numbers $|\mu|$.

\subsection{Upper and lower states perturbation}

To demonstrate the effect of dispersion modulation on the cavity dynamics for all comb modes, we provide split-step simulations~\cite{AGRAWAL201927} of Eq.~\ref{eq:LLE1}, shown in Fig.~\ref{fig:cw_dispersion}. We first investigate the effect of phase-matching on the noise transduction properties of the cavity, in the \emph{absence} of any coherent structure formation to avoid the effect of conventional modulation (Turing) instability~\cite{Conforti2014Modulational}, we simulate the case of a low average normal dispersion ($d_2^{(0)}=0.0027$). 
Given a CW driving strength of $f^2=10$, typical for the generation of solitons or switching waves, the Kerr nonlinear cavity possesses a bistable condition across a range of detuning to approximately $\zeta_0<10$, with an upper and lower CW-state solution $\Psi_H$ and $\Psi_L$, respectively (solutions of which in appendix). In Fig.~\ref{fig:cw_dispersion}(a,c), corresponding to the respective lower and upper states, we show the long-term response to small noise continuously placed on each comb mode, revealed by taking the 2-D Fourier transform of the output fast-time domain field recorded over a large number of roundtrips, with a numerical integration step smaller than the cavity roundtrip. Such a figure is henceforth referred to as the ``nonlinear dispersion relation'' (NDR), used in optics and hydrodynamics to describe complex nonlinear systems~\cite{Leisman2019EDR, Tikan2021EmergentDimer, tikan2021nonlinear}. It can be expressed as follows:
\begin{equation}\label{eq:NDR}
  \mathrm{NDR}(\Omega,\mu) = \frac{1}{\sqrt{N_t N}}\sum_{\ell,k} \Psi_{\ell,k} e^{i ( \Omega t_k - 2 \pi \mu \ell /N )},
\end{equation} 
here $\Omega$ is a slow frequency, $t_k = \Delta t k$ with $\Delta t = T_l/N_t$ time-step, $T_l$ is simulation time with $N_t$ number of discretization points.
Figure~\ref{fig:cw_dispersion} presents power spectral density (PSD) of the NDR over Floquet index $n$ vs. $\mu$, which are respectively proportional to the `slow' frequency $\Omega$ counted in $D_1$ and longitudinal mode (comb) index.
 
Each state carries conjugate pairs of resonant radiation conditions related to the dispersion operator \cite{milian_soliton_2014, he_dynamics_2016}, originating from the two states $\Psi_H$ and $\Psi_L$: $\zeta_{H\pm}$ and $\zeta_{L\pm}$ respectively (see appendix).
Our simulations reveal that the noise within the cavity forms a prominent resonance curve that follows one of those dispersion relations, depending on which bistable state the field is in. We note the conjugated $\zeta_+$ relation appears weakly only on the upper state (depicted as $\zeta_{H+}$ in Fig~.\ref{fig:cw_dispersion}(c)) as a result of sufficient FWM with $\zeta_{H-}$. 
In Fig.~\ref{fig:cw_dispersion}(b,d), we see how rapid dispersion frequency-modulation causes this radiation condition to carry sidebands, hence referred to as Floquet bands, spaced along the $n$ axis spaced by $d_1$ ($n=\zeta/d_1$). For this simulated example, $d_1/2\pi=8$ corresponding to 8 times the photon lifetime frequency. For simplicity, we consider an example of $\tilde{d}^{(1)}_{2} = \tilde{d}^{(-1)}_{2} = \Delta/2$ that corresponds to cosine modulation of the dispersion. In this case, the linear term in Eq.~(\ref{eq:2D_eq}) couples $n$ and $n\pm1$ frequency modes with a coupling strength $\Delta\mu^2/2$. This effect is similar to sideband generation in electro-optic modulation, in which here efficiency increases with the mode number $\mu$.  Strikingly, \emph{both} upper-state Floquet bands $\zeta_{H-}$ and $\zeta_{H+}$ are affected by the modulation in a similar way, as shown in Fig.~\ref{fig:cw_dispersion}(d).

\begin{figure*} [t]
\centering
\includegraphics{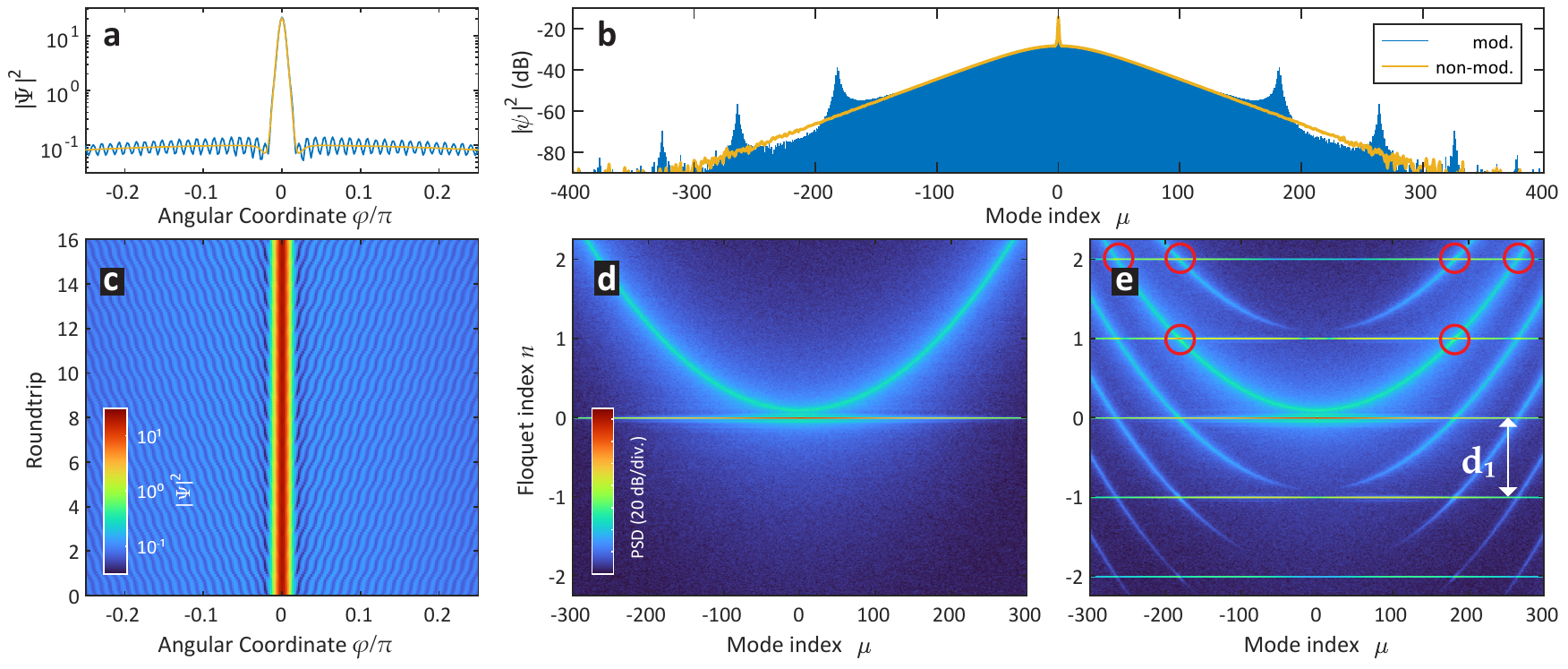}
\caption{\textbf{Dissipative Kerr soliton (DKS) simulation in CW-driven modulated cavity}. For $f^2=10$, $\zeta_0=10$, $\Delta=0.7 d_2^{(0)}$, $d_1/2\pi=16$, and $d_2^{(0)}=-0.0027$. (a) DKS field in cavity angular coordinate with (blue) and without (yellow) modulation. (b) Corresponding fast frequency power spectrum. (c) Spatiotemporal diagram of the modulated DKS propagating over resonator roundtrips (slow time). (d,e) DKS nonlinear dispersion relations, obtained by taken $\mathcal{F}[\;]$ over both dimensions of the spatiotemporal diagram for non-modulated and modulated cases, respectively. Red circles show higher-order dispersive wave positions.}
\label{fig:soliton_case_1}
\end{figure*}

\subsection{Faraday instability}
To reveal the emergence of Faraday instability (FI), we modify the simulations presented in Fig.~\ref{fig:cw_dispersion}(d) entering the range of parameters corresponding to the unstable regime. The above-mentioned simplification allows us to further develop an analytical derivation and analyze the linear stability of the system. We use a conventional stability analysis approach~\cite{Chembo2013Spatiotemporal}, but in the new 2D FWM setting created by Eq.~\ref{eq:2D_eq} assuming FSR$/2$ periodic dynamics of the field. As a result, we obtain that comb indices corresponding to the maximum FI gain can be approximated by the following expression (see Appendix~\ref{sec:Appendix_MI}): 
\begin{equation}\label{eq:mu_max_gain}
    \mu \approx \pm \sqrt{\frac{d_1}{2\sqrt{\big(d_2^{(0)}\big)^2-\Delta^2}}} \approx \pm \sqrt{\frac{D_1}{D_2^{(0)}}}.
\end{equation}

Equation~\ref{eq:mu_max_gain} reveals that in the normal dispersion regime, it is possible to observe formation of sidebands with frequency offset of FSR$/2$ from the pump which is also highlighted in Fig.~\ref{fig:FI_CW}.
Fig.~\ref{fig:FI_CW}(a-e) show the result of the numerical simulations.
The considered range of parameters correspond to period-doubling dynamics in the resonator with a $2T$ oscillation period, similar to results of Ref.~\cite{Bessin2019Real}. Fig.~\ref{fig:FI_CW}(a,b) show a corresponding spatiotemporal diagram and its cross-sections at two states separated by $T$. The NDR shown in Fig.~\ref{fig:FI_CW}(d) shows the 2D nature of the FWM pathways implying that pump photons can be transferred in the 2D frequency space changing both $\mu$ and n indexes. The maximum FI gain (see Fig.~\ref{fig:FI_CW}(e) for the full FI gain diagram) is placed at the modes corresponding to a $d_1/2$ spacing between the $\zeta_{H+}$ and $\zeta_{H-}$ Floquet bands.

\begin{figure*}[t]
	\centering
	\includegraphics[width=\linewidth]{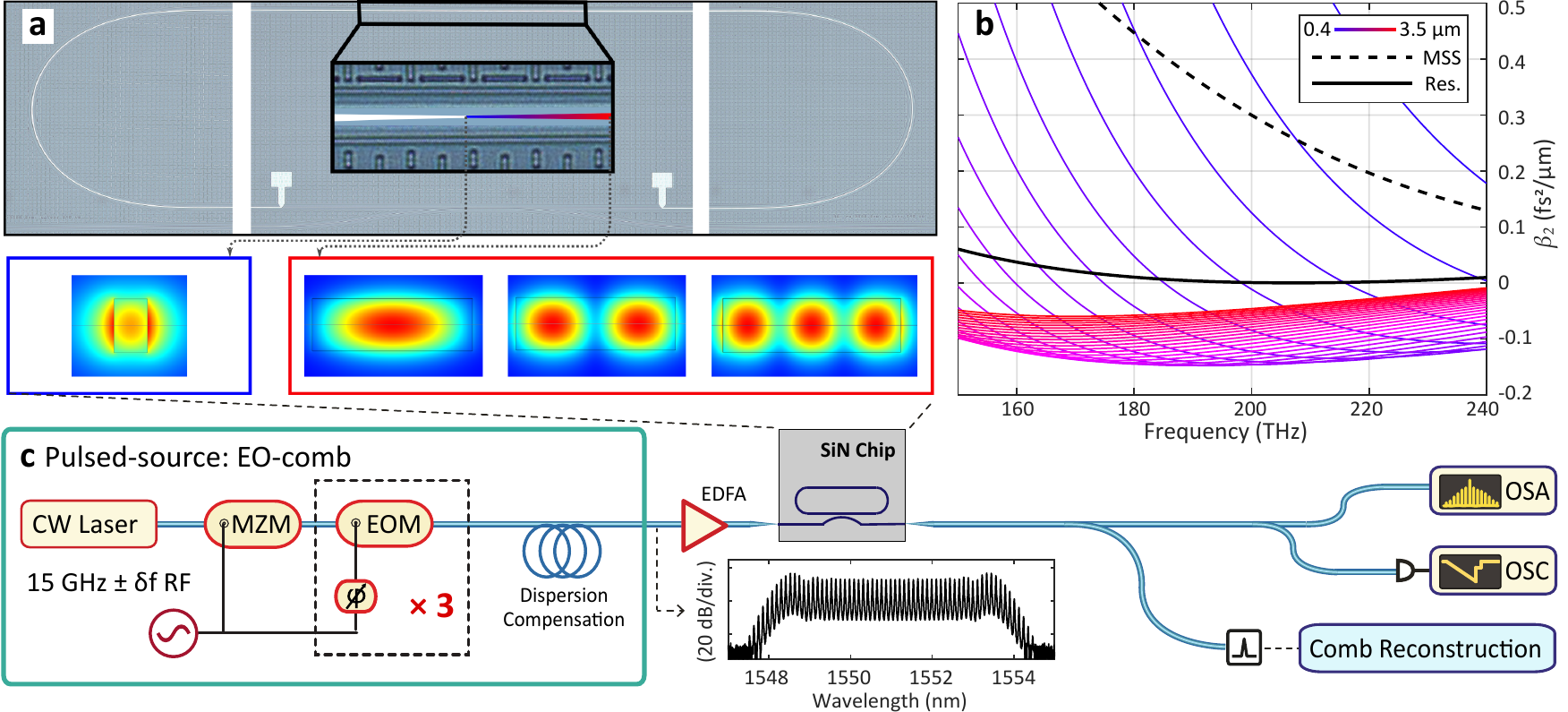}
	\caption{\textbf{\SiN photonic chip and experimental setup.} (a) Microscope image of the Si$_3$N$_4$ racetrack microresonator having 15 GHz free spectral range. The inset shows the mode suppression section (MSS). The variation of the waveguide width is highlighted by the color gradient. (b) The variation of the dispersion over the MSS for changing waveguide width (height = 820 $\upmu$m), the aggregate dispersion of the whole MSS, and aggregate dispersion for the whole resonator. (c) Experimental chip pumping scheme, featuring the EO-comb as a pulsed-source. The input pulse train is coupled into and out of the microresonator chip via lensed fibers. Bottom-inset: Spectrum of the 15 GHz EO-comb before amplification.  MZM: Mach-Zehnder modulator, EOM: electro-optic modulator, EDFA: erbium-doped fiber amplifier, OSA: optical spectrum analyzer, OSC: oscilloscope. See supplementary information for description of ``comb reconstruction'' section.}
	\label{fig:mod_concept1}
\end{figure*}

\section{Anomalous dispersion case}
\subsection{Numerical analysis}

First, we revisit the effect of the perturbation on the bright DKS formed in the anomalous dispersion cavity. The exact distribution of the modulation along the cavity can affect the position or the amplitude of the instability gain. However, this does not change overall dynamical features~\cite{Conforti2014Modulational}, which allows us to use the simplest cosine-modulated cavity. Although the Faraday instability gain can be positive, in this case~\cite{Copie2016Competing}, the CW solution on the upper branch is Turing unstable \cite{haelterman_dissipative_1992} and, in the considered range of parameters, leads to DKS generation. Therefore, dispersion modulation here acts as a roundtrip-periodic perturbation to a stable DKS state. This provides a photon transferring mechanism -- quasi-phase matching -- resulting in power enhancement in certain modes placed at the same frequency grid as the soliton line.  They modify the solitonic spectrum with Fano-shaped sidebands (also known as Kelly sidebands)~\cite{Nielsen2018Invited}, to which we refer to as here as higher-order dispersive waves (HDW).

To illustrate this, we repeat numerical solutions to Eq.~(\ref{eq:LLE1}). In Fig.~\ref{fig:soliton_case_1}(a,b,d), we recall the conventional (unperturbed) DKS features. Fig.~\ref{fig:soliton_case_1}(a) (yellow) shows a cross-section which contains a secant hyperbolic profile on a CW background. The frequency domain of this is shown in Fig.~\ref{fig:soliton_case_1}(b). Taking the 2D Fourier transform of the spatiotemporal data, we obtain NDR (described above), in Fig.~\ref{fig:soliton_case_1}(d). The NDR of a single unperturbed soliton has two components: a soliton line, and the dispersive resonance curve $\zeta_{L-}$, which is seeded mainly by the cavity noise, and here is approximately equal to the cold-cavity dispersion operator. The detuning of the DKS from the cold cavity resonance $\zeta_0$ is given by the gap between the soliton line and the $\zeta_{L-}$ curve. Crucially, even though the cavity dispersion wraps over the FSR line $n=1$ more than once, \emph{no dispersive wave} is created in this case since the nonlinear photon transfer is forbidden by the momentum conservation law~\cite{Huang2017Quasi}.

When periodic dispersion modulation is introduced, we observe a different picture (Fig.~\ref{fig:soliton_case_1}(a-c,e)). The spatiotemporal diagram (c) reveals that DKS starts to radiate dispersive waves to the cavity, depicted by wavy lines emanating  from the DKS and overlapping every roundtrip. They appear as a CW background modulation shown in (a), and in (b) are seen to be several HDW on the spectral wings (ie. Kelly sidebands). The NDR presented in Fig.~\ref{fig:soliton_case_1}(e) reveals that the HDW originate from the intersection between the soliton line and the FSR-folded Floquet bands. This interaction is enabled by the periodic modulation which couples neighboring modes, appearing in the NDR as copies of the soliton line and dispersive Floquet bands repeated at every FSR. In this way, the momentum conservation law can be satisfied which leads to efficient nonlinear photon transfer at the intersection points, forming the HDW in the spectrum. An extended 3D perspective and a view of the dispersion-folded space can be found in the supplementary information

\subsection{Experimental set-up}

The experimental setup used for all the experimental results of this work is presented in Fig. \ref{fig:mod_concept1}. The resonators of choice (one pictured in Fig.~\ref{fig:mod_concept1}(a)) are based on the photonic \SiN waveguide platform, fabricated using the \emph{Damascene process}~\cite{Liu2021High-yieldCircuits}, and have an FSR of 15 GHz. While the ring resonator in the theoretical model is assumed to have a sinusiodally varying second-order dispersion $D_2$, the real resonators are more complex. The original motivation for inserting a single mode section into the resonators was to suppress mixing between higher order transverse modes, that can lead to single mode dispersive waves. However, as detailed below we found that this leads to new dynamics. This was achieved with the use of a higher-order mode suppression section (MSS), a momentary length of the resonator where the waveguide tapers from its main waveguide width down to 0.4 $\upmu$m~\cite{kordts_higher_2016}, a width where only the fundamental mode may propagate without strong loss, shown in Fig. \ref{fig:mod_concept1}(a). 
In Fig. \ref{fig:mod_concept1}(b), we plot numerically calculated waveguide dispersion values $\beta_2$ for several discrete waveguide cross-sections along the MSS, showing how the dispersion transitions from weakly anomalous up to strongly normal. We can retrieve the \emph{average} roundtrip dispersion $\beta_2^{(0)}$ through a weighted sum of $\beta_2$ for each waveguide width according to their relative length. In this example resonator design, the final aggregate dispersion calculation comes out close to zero, since the MSS accounts for 13\% of the total cavity length. This also means the dispersion modulation duty-cycle is not pure sinusoidal, but rather more pulse-like with many more coupling harmonics in the longitudinal $n$-space. The final resonator dispersion value is found through 
\begin{equation}
	D_2^{(0)}=-\beta_2^{(0)}LD_1^3/2\pi
\end{equation} 
Resonators fabricated with this architecture for the experiment included a range of average dispersion values from anomalous to normal, providing the results for both regimes in this work.

\begin{figure*}[t]
	\centering
	\includegraphics[width=\linewidth]{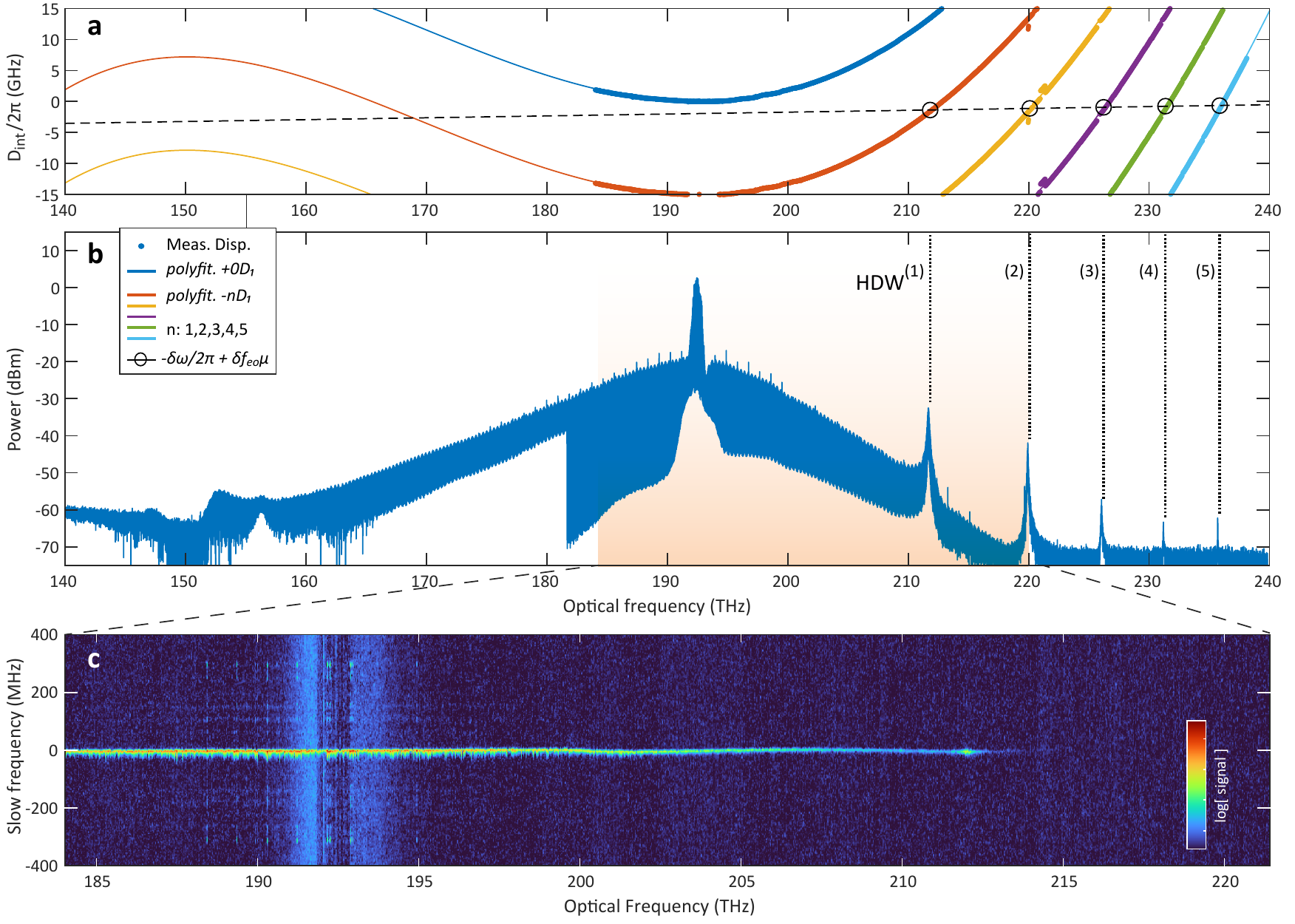} 
	\caption{\textbf{Dissipative soliton with higher-order dispersive waves experiment.}  (a) Measured integrated dispersion profile overlaid against polynomial fit, including all tranches separated by the FSR $D_1$. Dashed line represents comb line array. (b) Measured single soliton spectrum in resonator device with the above dispersion profile. Observed higher-order DW on high frequency side matched with points on the corresponding dispersion curves. (c) Experimental comb reconstruction measurement of the soliton comb in the frequency range 184-220 THz.}
	\label{fig:soliton_exp1}
\end{figure*}

The pumping and measurement setup itself is presented in Fig. \ref{fig:mod_concept1}(c). In order to reach the required driving powers for the regime where the Floquet dynamics due to the presence of dispersion bands can be accessed and to ensure the generation of a single dissipative structure, pulse-driving using an electro-optic comb is employed (EO-comb)~\cite{fujiwara_optical_2003} as it has been done previously for experiments in resonators with a similar GHz-domain FSR \cite{Obrzud2017TemporalPulses, Anderson2021PhotonicSolitons, xu_frequency_2021}. It yields a pulse train of pulses approximately 1.4 ps in duration with an RF controllable repetition rate $f_\mathrm{eo}$ set near 15 GHz (see appendix for further).  

\subsection{Experimental verification}
First, we verify the excitation of higher order dispersive waves in strongly pulse-driven resonators, that feature periodic dispersion as shown in Fig.~\ref{fig:soliton_exp1}. The resonator device used here, R1, has a relatively cubic dispersion profile with coefficients $D_{1,2,3,4}/2\pi=$ 15.06 GHz, 14.3 kHz, 6.59 Hz, and -3.84 mHz respectively, with intrinsic loss and coupling $\kappa_{0,\mathrm{ex}}/2\pi =$ 30 kHz and 230 kHz respectively. The high value of $\kappa_\mathrm{ex}$ was used in order to maximize the output power of the soliton spectrum, with a given average power of 720 mW entering the chip waveguide. After the soliton spectrum was generated stably, by tuning the EO-comb centre frequency across resonance into the bistability region~\cite{Herr2014Temporal}, there appeared several Kelly-like sidebands, or HDW, on the short-wavelength side of the spectrum. Remarkably, these higher order dispersive waves were observed up to 5th order and appeared spectrally highly distinct from the DKS. Extrapolating the DKS envelope, we observe that the HDW exhibits more than 25 dB power compared to the smooth single DKS case.
Fig. \ref{fig:soliton_exp1}(a) shows the integrated dispersion profile of this microresonator measured separately~\cite{liu_frequency-comb-assisted_2016}, overlaid with its fourth-order polynomial fitting $D_\mathrm{int} =\sum_{k=2}^{4} D_k\mu^k/k!$, and 5 additional orders of $D_\mathrm{int}$ separated negatively by $nD_1$. In Fig. \ref{fig:soliton_exp1}(b) the final soliton spectrum is plotted featuring HDW up to the fifth order. 

\begin{figure*} [t]
	\centering
	\includegraphics{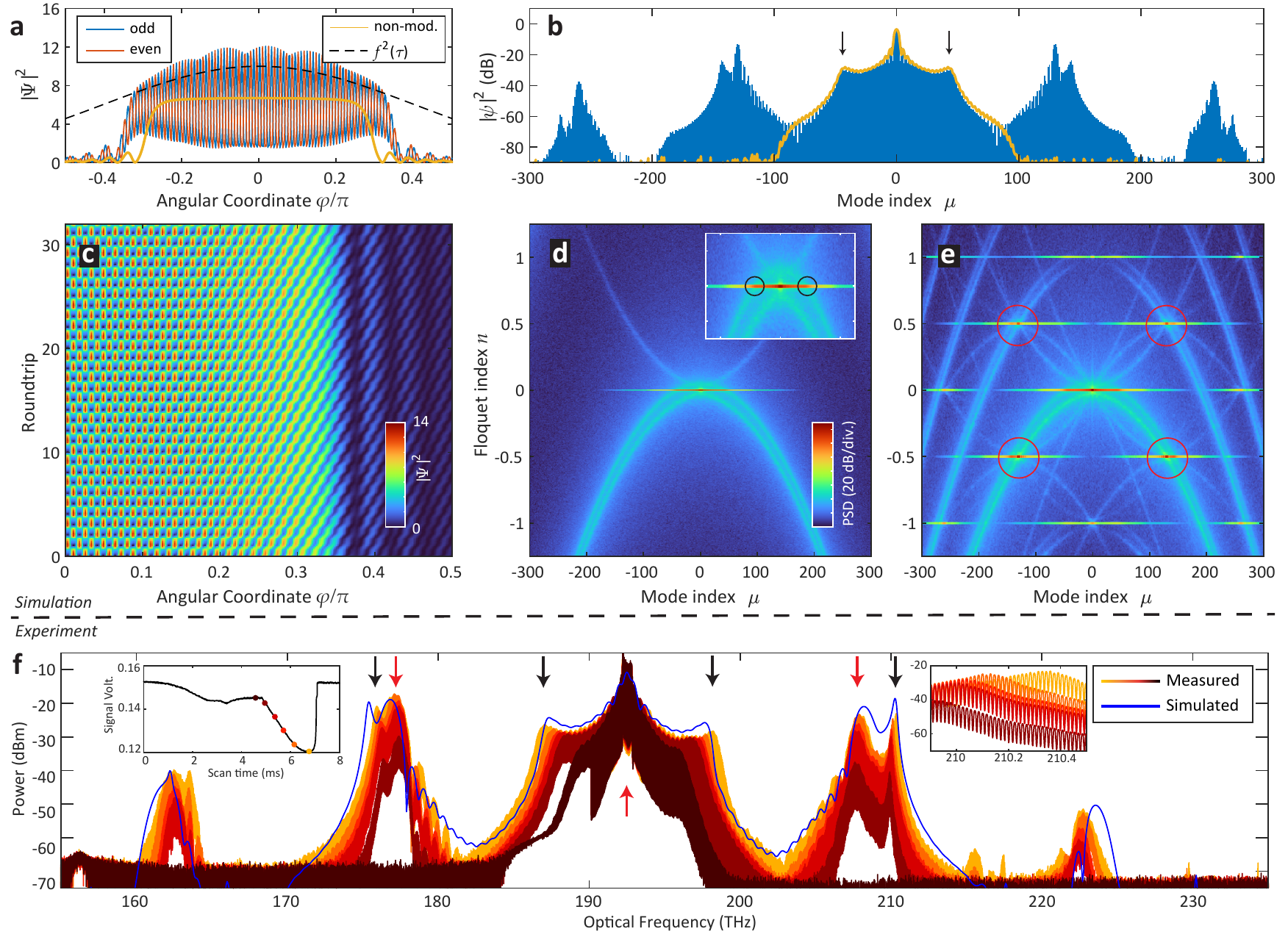}
	\caption{\textbf{Switching wave (SW) solution in pulse-driven modulated cavity} For $f^2=10$, $\zeta_0=6$, $\Delta=0.7 d_2^{(0)}$, $d_1/2\pi=16$, and $d_2^{(0)}=0.0027$. (a) SW field in cavity angular coordinate with (blue) and without (yellow) modulation. (b) Corresponding power spectrum. (c) Spatiotemporal diagram of an SW in the modulated cavity. (d,e) SW nonlinear dispersion relation, obtained by taken $\mathcal{F}[\;]$ over both dimensions of the spatiotemporal diagram for non-modulated and modulated cases, respectively. Black circles show the intersections of a SW line and dispersive parabola resulting in the canonical SW spectral profile. Red circles indicate FI-originated satellites. (f) Measured optical spectra of switching wave affected by FI at different detunings (red-orange scale). Simulated spectral envelope overlaid (blue). Left inset: measured cavity power out during scan of laser detuning over resonance. Red/orange points mark locations of measured spectra. Right inset: zoom-in of comb lines at outer dispersive wave. Black and red arrows correspond to circles from panels (d) and (e).}
	\label{fig:SW_case_1}
\end{figure*}

Next, we compare the location of the observed HDW with the theoretical predictions. By taking the frequencies of each HDW in Fig. \ref{fig:soliton_exp1}(b) and projecting them on to each Floquet order of \emph{folded} integrated dispersion operators in Fig. \ref{fig:soliton_exp1}(a) (black circles), we retrieve a linear frequency comb grid (black dash). This tilted line gives us the soliton comb relative frequency grid whose repetition rate (comb line spacing) is controlled by the injected EO-comb pulse-train, desynchronized from the cavity FSR slightly by $f_\mathrm{eo}=D_1/2\pi+\delta\!f_\mathrm{eo}$ \cite{Obrzud2017TemporalPulses}.
We directly confirm the comb integrity using the Kerr comb reconstruction technique~\cite{herr2012universal}, which allows us to experimentally obtain a direct measurement of the NDR (See Appendix for measurement details). The resulting image is shown in Fig. \ref{fig:soliton_case_1}(c), measured across the bandwidth available to us of $\sim$184-240 THz; although, only the first-order dispersive wave had a sufficient signal-to-noise ratio to be captured. The image shows that every comb line of this spectrum lies on a straight grid spaced by 15.05907 GHz, as sampled in this plot, which did exactly equal the experimentally set EO-comb frequency $f_\mathrm{eo}$. In particular, the first HDW seen at 211 THz lies on the very same grid.

It should be pointed out that the soliton spectrum does not match prediction on the long-wavelength side. According to the above dispersion plot, we should see likely two more first-order HDW as the comb grid crosses the $D_{L-}-D_1$ operator ($\approx D_\mathrm{int}-D_1$, see appendix) twice at 169 THz and later below 140 THz. Instead, we see these two features at 148 and 154 THz indicating the polynomial fitting is inaccurate in this region due the lack of measured dispersion values beneath 180 THz. 

\section{Normal dispersion case}

\subsection{Numerical analysis}

Next, we discuss the Floquet dynamics of coherent structures in normal dispersion resonators. In this case the pulsed pumping, used in our experiments to achieve high peak powers, plays another major role in stabilizing SW structures that appear in the resonator. SW usually have a non-zero relative group velocity that depends on the driving amplitude $f^2$. However, there is a particular value of $f^2$ that corresponds to a stationary SW pattern called a Maxwell point~\cite{Parra-Rivas2016Origin,anderson2020zero}. Pulse pumping leads to an intracavity power gradient that depends on the intra-resonator coordinate $\varphi$, therefore the edges of a SW lock to the part of the pump corresponding to the Maxwell point. The dispersion modulation, in this case, leads to the Faraday instability dynamics including the effect of period-doubling\cite{Bessin2019Real}, resulting in the generation of widely spaced sidebands that originate from the two-dimensional FWM process.

We provide numerical simulations of the LLE Eq.~\ref{eq:LLE1} and compare homogeneous and modulated cavity cases.
Fig.~\ref{fig:SW_case_1}(a-b,d) display the ideal SW generated in a synchronously-pumped resonator. %The spatiotemporal diagram (Fig.~\ref{fig:SW_case_1}(a)) shows half of the intracavity field dynamics since there is a mirror symmetry imposed by the pump at $\varphi = 0$.
Fig.~\ref{fig:SW_case_1}(a) (yellow) shows a typical platicon profile of a rectangular pulse with oscillating tails. Fig.~\ref{fig:SW_case_1}(b) is the corresponding spectrum showing the flat-top spectral profile. The NDR plotted in Fig.~\ref{fig:SW_case_1}(d), particularly the inset panel, clearly reveals the origin of such a spectrum. Since both branches of the bi-stable resonance $\Psi_L$ and $\Psi_H$ are involved in the SW formation, we observe all dispersive resonances $\zeta_{L-}$, $\zeta_{H-}$, and $\zeta_{H+}$ on the NDR ($\zeta_{L+}$ is again too weak to appear). The $\zeta_{H-}$ curve originating from the upper state $\Psi_H$ (top of the SW) acquires an additional phase shift due to the  Kerr nonlinearity and therefore is shifted down to lower frequencies relative to $\zeta_{L-}$. A coherent structure corresponding to the rising and falling edges of the SW acquires a smaller Kerr shift, and therefore caresses the curve $\zeta_{L-}$. Such a crossing implies the phase-matching between the states and leads to enhanced power at the crossing modes. This creates the flattened spectral profile of the SW \cite{xu_frequency_2021, anderson2020zero}.

When the SW is generated in the modulated cavity, a strong influence of the Faraday instability (FI) is observed. Performing numerical simulation with the same parameters as in the soliton section, but with the sign of $d_2^{(0)}$ reversed, we observe \emph{period-doubling} dynamics in  Fig.~\ref{fig:SW_case_1}(a,c).  The top part of the SW is patterned with periodic structures, similar to that observed in the CW case (Fig.~\ref{fig:FI_CW}), that are direct evidence of FI. Each roundtrip, the patterned profile exhibits a $\pi$ phase flip as reported in Ref.~\cite{coen_modulational_1997, Bessin2019Real}. The power profiles $|\Psi (\varphi,t)|^2$ at two stages of the period-doubling dynamics are displayed in Fig.~\ref{fig:SW_case_1}(a). Corresponding spectral plot in Fig.~\ref{fig:SW_case_1}(b) shows the appearance of characteristic \emph{double-peaked} sideband spectra substantially extending the unperturbed SW spectrum. In a microresonator environment, these sidebands manifest as \emph{satellite combs} drawing energy from the central comb.

Besides the conclusion from the analysis that predicts the position of the modes having maximum FI gain, there is an empirical understanding of the process that can be formulated by analyzing the NDR plot shown in Fig.~\ref{fig:SW_case_1}(e). In the normal dispersion case, conventional modulation instability does not affect the upper state of the bi-stable resonance. Because of the high power, the conjugated upper state dispersive curve $\zeta_{H+}$ becomes visible on the NDR. Both curves are experiencing modulation resulting in the FSR-spaced Floquet bands. At the location in the $n$ vs. $\mu$ space where $\zeta_{H-}$ and $\zeta_{H+}$ cross, we observe the formation of satellites. Due to the apparent mirror symmetry between $\zeta_{H-}$ and $\zeta_{H+}$, the intersection occurs at $\pm$FSR/2 in the slow frequency dimension. As explained above, this process can be seen as a two-dimensional FWM process, providing photon transfer from the pump to the sidebands having an $\pm$FSR/2 offset. The double-peak structure of subcombs can be readily explained with the NDR. This double-feature was in fact seen in numerical simulations shown in \emph{Staliunas et al.}~\cite{staliunas_parametric_2013} but was not discussed there in further detail. The coherent sub-comb line formed around the unstable mode, sourced from FI, crosses both $\zeta_{H-}$ and $\zeta_{L-}$ simultaneously, resulting in two peaks, the spacing of which corresponds to the separation between these dispersive curves. A larger 3D perspective of Fig.~\ref{fig:SW_case_1}(e), looking closely at this feature, can be found in the supplementary information.

\begin{figure*}[t]
    \includegraphics[width=\linewidth]{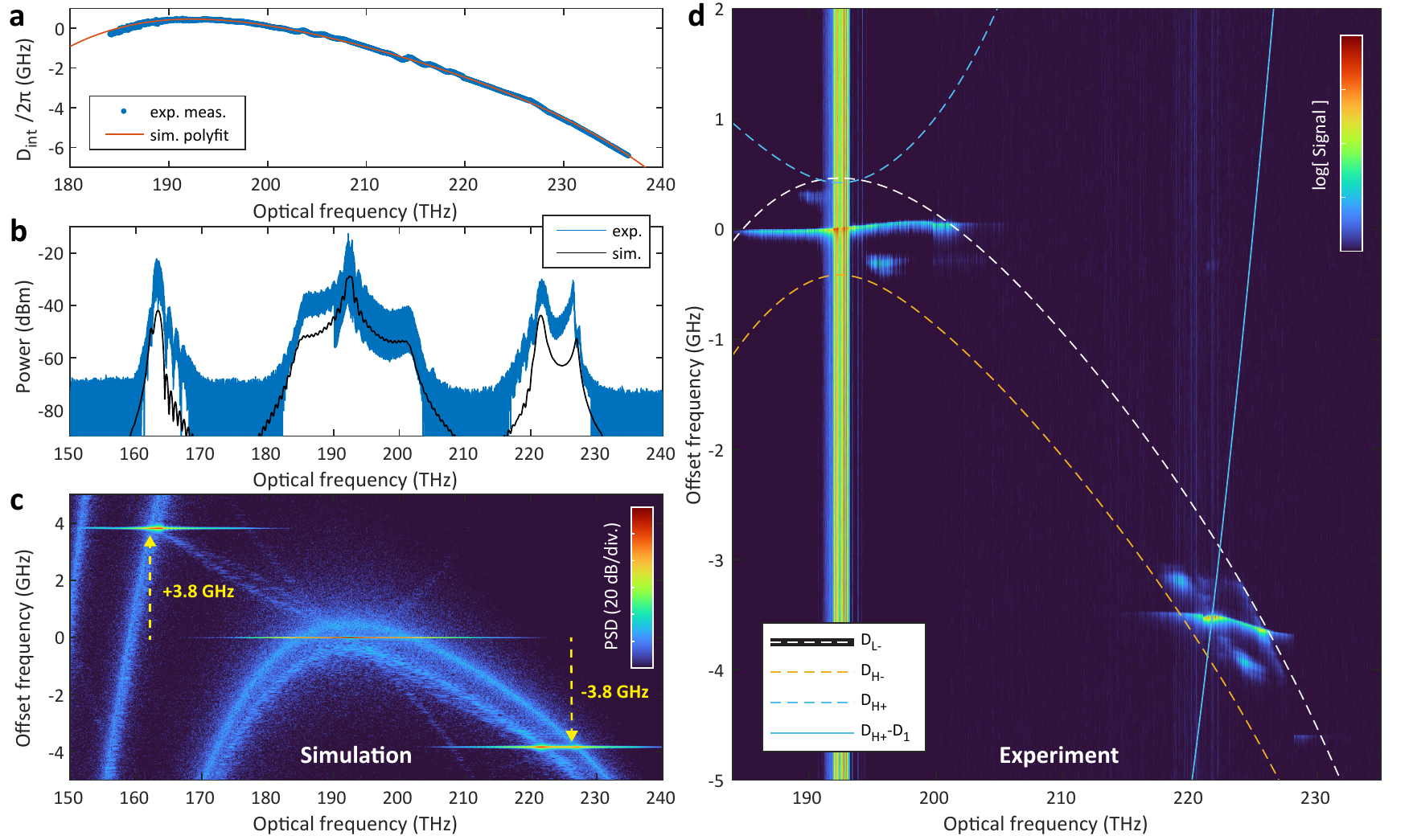}
    \caption{\textbf{Phase-matching for broadband satellite combs.} (a) Experimentally measured integrated dispersion for resonator R3, with fourth-order polynomial fit, used in the following simulation. (b) Experimentally measured spectrum of switching wave with satellite combs, overlaid with simulation result, offset by -15 dB. (c) Nonlinear dispersion relation of the simulation above, showing upper and lower dispersion curves and their higher orders. (d) Experimentally measured comb reconstruction, the equivalent of the image in (c), showing the central comb and short wavelength satellite located near the predicted phase-matched frequency offset. Overlaid are the analytical dispersion curves used from the simulation. The strong band across 192 THz is excess ASE from the pump spectrum.} 
    \label{fig:SWFI_wide_static}
\end{figure*}

\begin{figure*}[t]
    \centering
    \includegraphics[width=\linewidth]{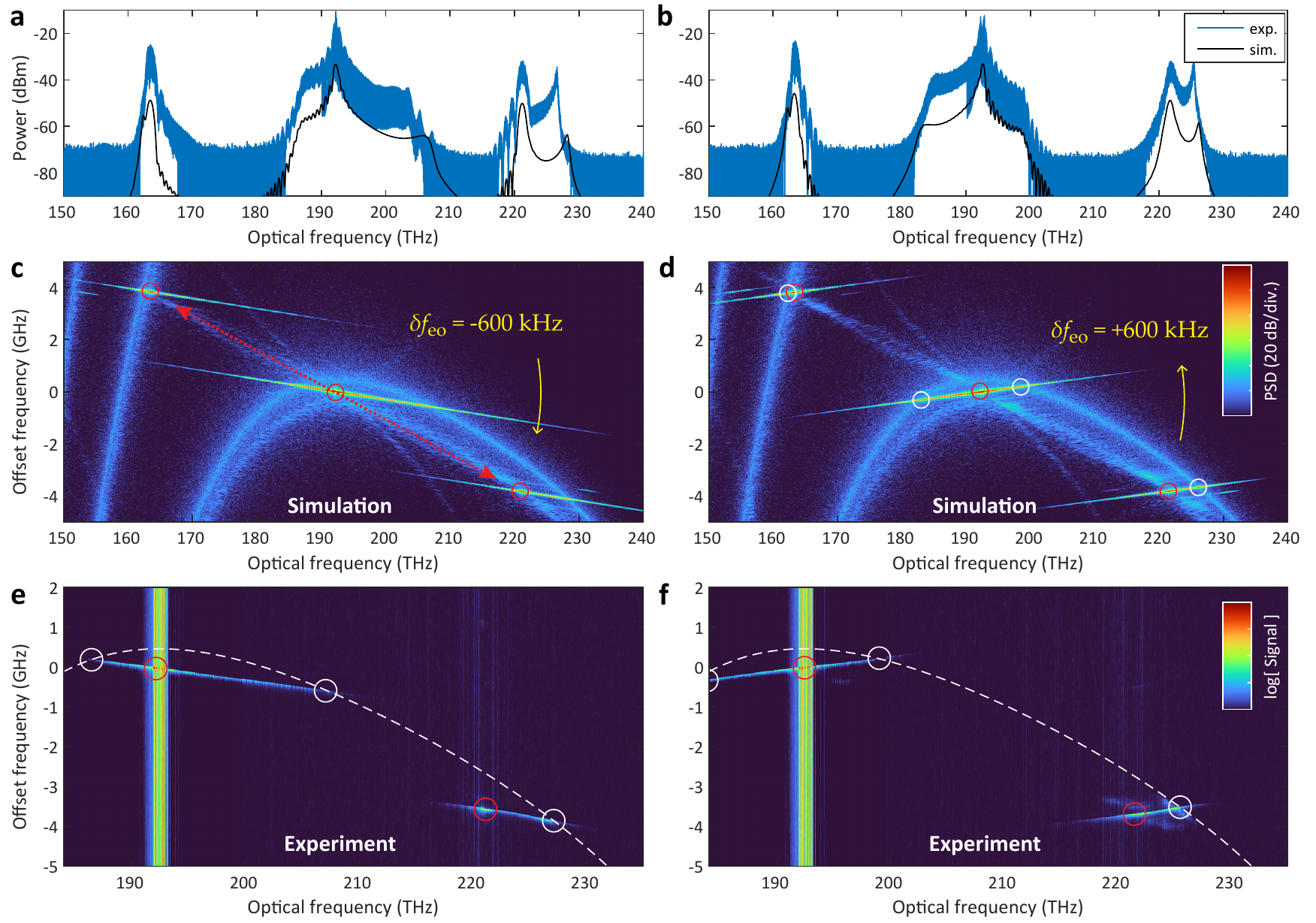}
    \caption{\textbf{Satellite comb repetition rate tuning and phase-matching.}. Two switching wave + satellite comb generation events where (a,c,d) $\delta\!f_\mathrm{eo}=+600$ kHz, and (b,d,f) $\delta\!f_\mathrm{eo}=-600$ kHz. (a,b) Spectrum, measured, and simulated using measured resonator parameters. Simulated is offset by -15 dB. (c,d) Nonlinear dispersion relation image for simulated case showing offset satellite comb spectra. (e,f) Measured comb reconstruction image. The strong band across 192 THz is excess ASE from the pump spectrum. Lower nonlinear dispersion curve shown in dashed-white. Faraday Instability gain frequency marked in red. Dispersive waves marked in white. }
    \label{fig:SWFI_wide_motion}
\end{figure*}

\subsection{Experimental verification}
We present experimental results of switching wave with satellite comb generation in dispersion-modulated resonators here, with the goal of verifying the coherence and comb-like nature of the spectrum as well as the validity of our modelling. For this, two example resonator devices are investigated: R2 and R3. R2 has relatively strong normal and symmetrical dispersion, with $D_{1,2,3,4}/2\pi=$ 15.32 GHz, -12.9 kHz, 5.36 Hz, and -2.39 mHz respectively, and $\kappa_{0,\mathrm{ex}}/2\pi =$ 30 kHz and 180 kHz respectively. R3, on the other hand, has a flatter and more imbalanced dispersion profile, with $D_{1,2,3,4}/2\pi=$ 15.31 GHz, -3.14 kHz, 3.35 Hz, and -2.49 mHz respectively, and $\kappa_{0,\mathrm{ex}}/2\pi =$ 30 kHz and 200 kHz respectively. All other experimental details including the driving and detection setup are the same as discussed above at Fig. \ref{fig:mod_concept1}. Further measurements of the resonator parameters can be found in the supplementary information. 

In Fig.~\ref{fig:SW_case_1}(f), we present satellite comb generation results in R2, with an average pump power of 230 mW. The left inset in this figure shows the detected power trace from the output of the cavity while the laser center-frequency is scanned across the cavity resonance from blue-detuned to red-detuned, showing different stages in the way the trace dips in power. The first dip and plateau corresponds to the initial build-up of power in the resonator, and the subsequent formation of the switching wave comb by wave-breaking~\cite{xu_frequency_2021, anderson2020zero}. The second dip, past halfway, marks the sudden growth of satellite combs. The reason the output power trace decreases in energy during this phase is due to the fact that the photodiode in use does not detect light above 1700 nm (below 176 THz). This indicates energy-transfer from the central comb to the satellite comb. As marked with red-to-orange dots, we stop the laser tuning at the several points here, and plot the spectra measured. Two large satellite combs are observed to rise quickly, \emph{above} even the plateau comb power of the core switching wave. Each satellite comb consists of two features. On the inner side closer to the core spectrum are the $\mathrm{sinc}(\;)$ profile-like spectra, marked with red arrows, that represent the origin of the FI pattern sourced from the `inside' of the switching wave, on its upper state, as shown previously in Fig. \ref{fig:SW_case_1}. These are equispaced about the core spectrum at $\pm$15.5 THz, due to the FI requirement, and appear to agree approximately with the prediction of Eq.~\ref{eq:mu_max_gain}. The second feature is the sharp hook-like dispersive wave on the outer side of the satellite combs, marked with black arrows. These mark the relative location of the lower-state dispersive resonance $D_{L-}$ (where $D_{L-}=\kappa\zeta_{L-}/2$ in dimensionless) with which the FI spectra interact, as also described above in Fig. \ref{fig:SW_case_1}. Using the independently measured loss and dispersion values of R2 given above, we simulated the very same experiment using the LLE with real resonator units. The final simulated spectrum when tuning into resonance is plotted in the same Fig.~\ref{fig:SW_case_1}(f) in blue, showing good agreement, particularly with respect to the satellite combs' `double-feature'. Further results from this simulation can be found in the supplementary information.

Moving on to further explore the phase-matching of these satellite combs more deeply, we present measurements and corresponding simulations of generated satellite combs in resonator device R3, shown in Fig. \ref{fig:SWFI_wide_static}. The experimental dispersion measurement is shown in Fig. \ref{fig:SWFI_wide_static}(a) and its fitting is used to obtain the dispersion parameters listed above, and carry out verification simulations. When tuning the laser to the point of strongest comb generation, we measure the spectrum shown in Fig. \ref{fig:SWFI_wide_static}(b), with average power coupled to the resonator of 60 mW, this time showing satellite comb generation further apart in comparison to Fig.~\ref{fig:SW_case_1}(f). Re-creating the same experiment in simulation creates the black spectral envelope, showing good agreement. In Fig. \ref{fig:SWFI_wide_static}(c), the NDR image for the simulation state is shown, revealing the coherence of the satellite combs and their precise relative position in offset frequency relative to the central switching wave comb, including the three nonlinear dispersion curves $D_{L-}$, $D_{H-}$, and $D_{H+}$ that determine this phase matching. According to this figure, the requirement for the FI gain to be equidistant with the pump, and located on the upper-state dispersion curve, causes the satellite combs to originate at 163 THz and 222 THz or $\pm$ 29.5 THz relative to the pump, with an offset frequency of $\pm$ 3.8 GHz respectively. In a perfectly symmetrical dispersion profile such as in Fig. \ref{fig:SW_case_1} or close to the experiment for R2 in Fig. Fig.~\ref{fig:SW_case_1}(f), this offset value would be at or near the FSR/2, 7.66 GHz. In R3, due to the positive $D_3$, the dispersion landscape has been pulled upwards in the positive optical frequency direction, resulting in this frequency offset value of 3.8 GHz.

To confirm our understanding, we perform comb reconstruction (described in the appendix), and plot the image in Fig. \ref{fig:SWFI_wide_static}(d). The data shown here corresponds to the same spectral measurement plotted in Fig. \ref{fig:SWFI_wide_static}(b). Plotted on top are the nonlinear dispersion curves calculated for the simulated parameters: the upper- and lower-state curves $D_{H-}$ and $D_{L-}$ respectively (dashed), as well as the conjugated upper-state curve $D_{H+}$ (dashed) and its FSR-shifted Floquet band (solid). In real units, $D_{\pm}=\kappa\zeta_{\pm}/2$.
In this image, we can see the central SW comb spanning 185-203 THz, and we verify that its spectral wings are bound by the lower-state dispersion curve $D_{L-}$. Crucially, we verify that the short wavelength satellite comb appears at an offset frequency of -3.5 GHz, almost where it is predicted to appear according to the simulated model at the intersection of $D_{H-}=D_{H+}-D_1$. The outer dispersive wave wing of the satellite also appears as this comb's intersection with the lower-state dispersion curve $D_{L-}$. 
Both combs appear to be coherent, with optical linewidths limited by the detection resolution of the probing laser ($\sim4$ MHz). We must acknowledge that the detected offset frequency of each comb line seems to drift up and down by several MHz across the whole measurement, particularly at the end of the satellite comb. This drift is likely due slow thermal shifts in the pumped resonator mode spectrum (which is not locked) occuring during the 40 s swept-laser scan. This varies from take to take. It is not indicative of the physical frequency comb. 

To complete our demonstration of the nature of the satellite comb offset frequency, and its relationship to the dispersion curves, the same series of graphs is presented in Fig. \ref{fig:SWFI_wide_motion}. This time, however, the frequency of the EO-comb $f_\mathrm{eo}$ ie. the imposed repetition rate of SW generation inside the cavity, is changed between two limits. In the above Fig. \ref{fig:SWFI_wide_static} results, $\delta\! f_\mathrm{eo}$ was set to 0 in simulation, and in experiment it was set to the point of maximum spectrum-symmetry in the wings of the SW comb, with $f_\mathrm{eo}=$ 15.30877 GHz. In Fig.\ref{fig:SWFI_wide_motion}(a,c,d) we downturn $f_\mathrm{eo}$ by $\delta\! f_\mathrm{eo}=-600$ kHz, and in Fig.\ref{fig:SWFI_wide_motion}(b,d,f) upturn by $\delta\! f_\mathrm{eo}=+600$ kHz. In each case, experimentally and in simulation, it is important to observe that the centre of FI gain stays \emph{fixed} at $\pm29.5$ THz in the optical axis, and at $\pm3.8$ GHz or $\pm3.5$ GHz offset frequency in simulation and experiment respectively. These points are highlighted by red lines and circles. While these points remain fixed, the repetition rate (comb line spacing) of the all combs exactly follows the imposed repetition rate, $f_\mathrm{rep}=f_\mathrm{eo}+\delta\! f_\mathrm{eo}$, causing the visual `tilt' in the combs shown in Fig. \ref{fig:SWFI_wide_motion}(c-f). Where the tilted comb spectra cross $D_{L-}$ is where the dispersive waves form, highlighted in white circles. By cross-comparing the NDR graphs with the optical spectra shown above in Fig. \ref{fig:SWFI_wide_motion}(a,b), one can see the direct origin of each comb feature. 

\section{Conclusion}
In this work we presented a comprehensive investigation of dissipative structure existence in dispersion-folded Kerr cavities, with a fundamental spatial modulation of the dispersion. The source of additional spectral features is shown directly through 2D analysis of four-wave mixing pathways, in simulation via the slow vs fast frequency NDR diagrams, and in experiment by the comb reconstruction measurement. 
For the dissipative soliton, we see directly the appearance of resonant radiation that we term as higher-order dispersive waves, ie. dispersive waves appearing quasi-phase matched to dispersive resonance curves (or Floquet bands) in FSR-folded space. This is termed as such in relation to conventional ‘zeroth-order’ dispersive waves that phase-match directly to the base level un-modulated dispersive resonance curve, such as dual-dispersive waves in resonators with quartic dispersion~\cite{Pfeiffer2017Octave}. 
For the switching wave structure, we showed direct generation of powerful satellite combs born of the Faraday Instability from the switching wave upper-state. These satellite combs are observed to be mutually locked in repetition rate with the core SW structure. However, the center frequency of each satellite comb is not determined by this repetition rate, and is instead set fixed by the cavity dispersion, forming at a point up to an FSR/2 offset from the core comb. 
We showed how the repetition rate is controllable by varying the injected pulse repetition rate.
Many of the cited works here instead predicted or observed similar structures arising as a result of the modulation of other parameters of the LLE, such as the nonlinear constant~\cite{staliunas_parametric_2013}, and the loss/coupling value\cite{coen_modulational_1997, Conforti2016} (via the \emph{Ikeda map} model as opposed to the LLE). This particular work dealt specifically with dispersion modulation.  All such processes can be analyzed using a similar method. 

In terms of applicable value, if assuming that engineering highly flat dispersion is difficult in a given waveguide platform, deliberate modulation of the dispersion instead will automatically generate many higher-order dispersive waves that extend a soliton microcomb spectrum significantly wider beyond the point where the body of the comb disappears into the optical noise floor. Doing so deliberately may extend an already broad soliton microcomb further to the point where it can become octave spanning and able to be $f$-$2f$ self-referenced~\cite{Spencer2018AnPhotonics}.
Similarly, and more dramatically, assuming a fixed dispersion profile, dispersion modulation is shown here to effortlessly enable a switching wave microcomb to be extended by multiple times its base level bandwidth. Such combs could in the future find use in spectroscopy~\cite{Picque2019Frequency}, telecommunications~\cite{fulop2018high}, astro-spectrometer calibration~\cite{Obrzud2019AAstrocomb}, and LiDAR~\cite{Riemensberger2020}. 
For these reasons, longitudinal parametric modulation of the resonator waveguide may become a resource for spectrally extending Kerr microcombs.

\section{Acknowledgments}

This work was supported by Contract No. D18AC00032 (DRINQS) from the Defense Advanced Research Projects Agency (DARPA). This material is based upon work supported by the Air Force Office of Scientific Research under award number FA9550-19-1-0250. This work was further supported by the European Union’s Horizon 2020 Programme for Research and Innovation under grant no. 812818 (Marie Skłodowska-Curie ETN MICROCOMB), and by the Swiss National Science Foundation under grant agreement 192293.
We also thank Alisa Davydova for aid in initial calculations, and Zheru Qiu for assistance in sample fabrication. 
All samples were fabricated in the Center of MicroNanoTechnology (CMi) at EPFL.
\appendix

\section{Analytical approach to the modulated cavity dynamics}

\subsection{Nonlinear dispersion operators}

The Kerr bistable solutions are found from the roots of the cubic equation of the LLE, at equilibrium where $\partial_t\Psi=0$, in the absence of any dispersion term:

\begin{equation}
	\label{eq:polynom3}
	(|\Psi|^2)^3-2\zeta_0(|\Psi|^2)^2+(\zeta_0^2+1)|\Psi|^2-f^2=0
\end{equation} 

and subsequently

\begin{equation}
	\label{eq:nonlin_res}
	\Psi_\mathrm{CW}=\frac{if}{|\Psi_\mathrm{CW}|^2-\zeta_0+i}
\end{equation} 

The resulting nonlinear dispersion operators are found from \cite{milian_soliton_2014} as follows:

\begin{align}
	\label{eq:radiation_law}
	\zeta_{L\pm}(\mu) &= -\delta d_1\mu +d_3\mu^3 \nonumber\\ &\mp\sqrt{ (\zeta_0  -d_2\mu^2-d_4\mu^4 -2|\Psi_L|^2)^2 -|\Psi_L|^4 } \\
	\zeta_{H\pm}(\mu) &= -\delta d_1\mu +d_3\mu^3 \nonumber\\ &\mp\sqrt{ (\zeta_0 -d_2\mu^2-d_4\mu^4 -2|\Psi_H|^2)^2 -|\Psi_H|^4 } 
\end{align}

including orders of dispersion and group velocity shift from $d_1$ to $d_4$. 

\subsection{Modulation instability analysis}\label{sec:Appendix_MI}
In case of harmonic modulation of the dispersion, Eq.~(\ref{eq:2D_eq}) indicates linear coupling between the amplitudes $\tilde{\psi}_{n \mu}$ and $\tilde{\psi}_{n\pm1 \mu}$, resulting in breathing of generated structures with period equal to the round-trip time. However, the presence of Kerr nonlinearity gives rise to period multiplication effects that occur due to Faraday instability. In this section we perform a modulation instability analysis for the period-doubled continous-wave solution.

We assume $\psi_0(\varphi,t)$ to be a solution of Eq.~(\ref{eq:LLE1}) with an unmodulated dispersion term. We are interested in dynamics of a small perturbation $\xi(\varphi, t)$ that obeys the linearized equation

\begin{align}
    &\frac{\partial \xi}{\partial t} = -(1+i \zeta_0 )\xi +i\big( d_2^{(0)} + \Delta\frac{e^{i d_1 t}+e^{-i d_1 t}}{2} \big)\frac{\partial^2}{\partial \varphi^2}\xi +\nonumber\\
&+i(2|\psi_0|^2\xi + \psi_0^2 \xi^*) + \Delta\frac{\partial^2 \psi_0}{\partial \varphi^2}\cos{d_1 t}.
\end{align}
First, we use the following ansatz $\xi = A(t)\exp{i\mu \varphi} + B^*(t)\exp{-i\mu\varphi}$, so the coupled equations for the amplitudes $A$ and $B$ read
\begin{align}
    \begin{cases}
        &\frac{\partial A}{\partial t} = -(1+i \zeta_0 )A -i\big( d_2^{(0)} + \Delta\frac{e^{i d_1 t}+e^{-i d_1 t}}{2} \big)\mu^2A +\nonumber\\
&+i(2|\psi_0|^2A + \psi_0^2 B) + \Delta\frac{\partial^2 \psi_0}{\partial \varphi^2}\cos{d_1 t},\\
        &\frac{\partial B}{\partial t} = -(1-i \zeta_0 )B +i\big( d_2^{(0)} + \Delta\frac{e^{i d_1 t}+e^{-i d_1 t}}{2} \big)\mu^2B +\nonumber\\
&-i(2|\psi_0|^2B + \psi_0^2 A) + \Delta\frac{\partial^2 \psi_0}{\partial \varphi^2}\cos{d_1 t}.
    \end{cases}
\end{align}
We continue, assuming $A = \alpha_+\exp{i d_1 t/2} + \alpha_-\exp{-i d_1 t/2},\, B= \beta_+\exp{i d_1 t/2} + \beta_-\exp{-i d_1 t/2}$, where the amplitudes obey 
\begin{equation}
    \frac{d}{dt}\mathbb{Y} = \mathbb{M}\mathbb{Y},
\end{equation}
where $\mathbb{Y}=[\alpha_+,\alpha_-,\beta_+,\beta_-]^T$ and the matrix

%\begin{widetext}
\begin{equation}
\mathbb{M}=
    \begin{pmatrix}
        y_{00} & -i \mu^2 \Delta/2 & i \psi_0^2 & 0 \\
        -i \mu^2 \Delta/2 & y_{11}& 0 & i \psi_0^2\\
        -i \psi_0^{*2} & 0 & y_{00}^* & i\mu^2 \Delta/2 \\
        0 & -i\psi_0^{*2} & i\mu^2\Delta/2 & y_{11}^*
    \end{pmatrix},
\end{equation}
%\end{widetext}
where $y_{00} = -(1+i \zeta_0) - id_1/2 - i\mu^2 d_2^{(0)} + 2i|\psi_0|^2$ and $y_{11} = y_{00}+id_1$. 
Real part of the eigenvalues of this matrix give the parametric gain of modulationally unstable solutions. Comb indices of these solutions can be approximated by the following formula 
\begin{equation}
    \mu = \pm \sqrt{\frac{d_1}{2\sqrt{(d_2^{(0)})^2-(\Delta/2)^2}}}
\end{equation}

\section{Numerical modeling}
To model the dynamics of the field envelope in periodically varying dispersion resonators, we numerically solved Eq.~(\ref{eq:LLE1}) in normalized units using well-known Split-Step method~\cite{AGRAWAL201927}. This was for the studies presented in Fig. \ref{fig:cw_dispersion},\ref{fig:FI_CW},\ref{fig:soliton_case_1},\ref{fig:SW_case_1}. For Fig. \ref{fig:SW_case_1}(f),\ref{fig:SWFI_wide_static},\ref{fig:SWFI_wide_motion}, the real experimental results were replicated in simulations using the system model with practical units:

\begin{align}
	\label{eq:LLE2}
	\frac{\partial A}{\partial t'} =\: & \mathcal{F}_\tau \Big[ i\big( \delta\omega +2\pi\mu\delta\!f_\mathrm{eo} +D_\mathrm{int}(t',\mu) \big) \tilde{A}_\mu \Big] \\ \nonumber - & \frac{\kappa}{2}A  + i\Gamma|A|^2A +f_p(\tau)\sqrt{\frac{D_1\kappa_\mathrm{ex}}{2\pi}P_0} 
\end{align}

acting on field (in $\sqrt{W}$) $A(t',\tau)$ over slow/laboratory time $t'$ and fast time $\tau$ in the co-moving frame of the intracavity field circulating at $D_1$, with frequency domain counterpart $\tilde{A}_\mu$ at discrete comb line indices $\mu$. The linear phase operators include the detuning $\delta\omega$, the desynchronization of the pulse-drive with the cavity field $\delta\!f_\mathrm{eo}$, and the dispersion operator that varies with time as 

\begin{equation}
	D_\mathrm{int}(t',\mu) = (1 +\Delta\cos(D_1 t'))\big[\frac{D_2}{2}\mu^2 +\frac{D_3}{6}\mu^3 +\frac{D_4}{24}\mu^4\big]
\end{equation}

The loss and coupling rates satisfy $\kappa=\kappa_0+\kappa_\mathrm{ex}$, and the nonlinear coupling parameter $\Gamma=D_1n_2\omega_0L/(2\pi cA_\mathrm{eff})$ for resonator length $L$, effective waveguide area $A_\mathrm{eff}$, and nonlinear refractive index $n_2$. The injected pulse profile has a peak power $P_0$. All parameters here are found experimentally and can be related to the dimensionless values by the following transformations: $t=t'\kappa$, \quad$\varphi=\tau D_1$, \quad$\Psi=A\sqrt{2\Gamma/\kappa}$, \quad$d_l=2D_l/(\kappa l!)$ for $l=$1-4, \quad$\zeta_0=2\delta\omega/\kappa$, \quad$f^2 = 4P_0\kappa_\mathrm{ex}\Gamma D_1/(\pi\kappa^3)$. The experimental nonlinear dispersion curves are related to the dimensionless curves by $\zeta_{\pm}(\mu)=2D_{\pm}(\mu)/\kappa$. All parameter values used to reconstruct the field dynamics are presented in the supplementary information.

\section{Experimental detail}
\subsection{Chip pumping}

The experimental setup for characterization of nonlinear frequency mixing and dissipative Kerr soliton generation combines the experimental setups used in Refs. \cite{herr2012universal} for Kerr comb reconstruction and \cite{Anderson2021PhotonicSolitons} for dissipative soliton generation and EO-comb control. A widely tuneable external cavity diode laser is passed through 3 fiber-coupled phase modulators and one amplitude modulator all synchronously driven by an RF synthesizer (Rhode \& Schwarz SMB100A) set to near the FSR. The 1.4 ps pulses are formed after traveling through approximately 275 m of SMF-28, then amplified in an erbium doped fiber amplifier, and coupled to the photonic chip through lens fibres with 2.4 dB insertion loss. 

\subsection{Comb reconstruction}
This technique is realized on the linear dispersion measurement tool~\cite{liu_frequency-comb-assisted_2016} reconfigured to a heterodyne optical spectrum analyzer by superimposing the output of the resonator with the scanning laser on a balanced photodetector. Kerr comb reconstruction provides a high spectral resolution (of the order of 4~MHz) and an extended dynamic range (enhanced by a multistage logarithmic amplifier - Analog Devices 8307) which allow us to obtain full spectral information about the generated comb state. The diagram of this measurement method is shown in the supplementary information.

\bibliography{biblio}

\end{document}